\documentclass[conference]{IEEEtran}
\usepackage{color}

\usepackage{graphicx}
\usepackage{multirow}
\usepackage{booktabs}
\usepackage[table]{xcolor}

\begin{document}
%
\title{Survey of Extended LEACH-Based Clustering Routing Protocols for Wireless Sensor Networks}

\author{\IEEEauthorblockN{M. Aslam, N. Javaid, A. Rahim, U. Nazir, A. Bibi, Z. A. Khan$^{\$}$\\}

        Department of Electrical Engineering, COMSATS\\ Institute of
        Information Technology, Islamabad, Pakistan. \\
        $^{\$}$Faculty of Engineering, Dalhousie University, Halifax, Canada.
        }
        
\maketitle

\begin{abstract}
An energy efficient routing protocol is the major concern infield of wireless sensor network.  In this survey paper we present some energy efficient hierarchal routing protocols, developed from conventional LEACH routing protocol. Main focus of our study is how these extended routing protocols work in order to increase the life time  and how quality routing protocol is improved for the wireless sensor network. Furthermore this paper also highlights some of the issues faced by LEACH and also explains how these issues are tackled by extended versions of LEACH. We compare the features and performance issues of each hierarchal routing protocol.\\

\end{abstract}

\IEEEpeerreviewmaketitle

\section{Introduction}
Recent advancement in electronics technology enabled designer to develop low cost, low power and small size sensors [1], [2]. Hundreds and thousands of these sensors are deployed in wireless sensor network according to the requirement of network application. Wireless sensor network (WSN) is one of the evolving technologies. Sensor nodes are able to monitor physical environment, compute and transmit this information to core network.These sensors can communicate to each other and also to some external Base station [8]. Wireless sensor network are used for both military and civil applications [5]. A wide-range of applications offered by WSN, some of these are environmental monitoring, industrial sensing, infrastructure protection, battlefield awareness and temperature sensing.\\

Routing is main challenge faced by wireless sensor network. Routing is complex in WSN due to dynamic nature of WSN, limited battery life, computational overhead, no conventional addressing scheme, self-organization and limited transmission range of sensor nodes [2], [3], and [4].. As sensor has limited battery and this battery cannot be replaced due to area of deployment, so the network lifetime depends upon sensors battery capacity. A Careful management of resources is needed to increase the lifetime of the wireless sensor network. Quality of routing protocols depends upon the amount of data ( actual data signal) successfully received by Base station from sensors nodes deployed in the network region.

Number of routing protocol have been proposed for wireless sensor network. Mainly these are three types of routing protocols\\
\begin{enumerate}
  \item Flat routing protocols
  \item Hierarchical routing protocols
  \item location based routing protocols
\end{enumerate}

The category of Hierarchical routing protocol is providing maximum energy efficient routing protocols [1], [2], [3], [4], [7], [8], [9], [10], [11], [12], [13]. Number of hierarchical routing protocol has been proposed. LEACH (Low Energy Adaptive Clustering Hierarchy) is considering as a basic energy efficient hierarchical routing protocol. Many protocols have been derived from LEACH with some modifications and applying advance routing techniques. This survey paper discus and compare few hierarchical routing protocols solar-aware LEACH (sLEACH), Multi-Hop LEACH, M-LEACH. These all are energy efficient routing protocols and provide quality enhancement to LEACH.\\
The rest of the paper is organized as follow. We discuss the LEACH, sLEACH, Multi-Hop LEACH and M-LEACH in next section. After that we will compare the features these selected hierarchical routing protocols and in section III. in IV section analytical comparison is given to elaborate the energy efficiency of routing protocols.Simulation results are discuss in section V. The last section concludes our comprehensive survey.
\section{HIERARCHICAL ROUTING PROTOCOLS}
In hierarchical routing protocols whole network is deviled into multiple clusters. one node in each cluster play leading rule. cluster-head is the only node that can communicate to Base station in clustering routing protocols.This significantly reduces the routing overhead of normal nodes because normal nodes have to transmit to cluster-head only [14], [1], [2], [3], [5], [7], [11], [12], [15]. Description of some hierarchical routing protocols is discuss in next subsections.
\subsection{LEACH (Low Energy Adaptive Clustering Hierarchy)}
LEACH is one of the first hierarchical routing Protocols used for wireless sensor networks to increase the life time of network. LEACH performs self-organizing and re-clustering functions for every round [1]. Sensor nodes organize themselves into clusters in LEACH routing protocol.  In every cluster one of the sensor node acts as cluster-head and remaining sensor nodes as member nodes of that cluster. Only cluster-head can directly communicate to sink and member nodes use cluster-head as intermediate router in case of communication to sink. Cluster-head collects the data from all the nodes, aggregate the data and route all meaningful compress information to Sink. Because of these additional responsibilities Cluster-head dissipates more energy and if it remains cluster-head permanently it will die quickly as happened in case of static clustering. LEACH tackles this problem by randomized rotation of cluster-head to save the battery of individual node [1], [2]. In this ways LEACH maximize life time of network nodes and also reduce the energy dissipation by compressing the date before transmitting to cluster-head.
LEACH routing protocol operations based on rounds, where each round normally consists of two phases. First is setup phase and second is steady state phase. In setup phase cluster-head and cluster are created. Whole network nodes are divided into multiple clusters. Some nodes elect themselves as a cluster-head independently from other nodes. These nodes elect themselves on behalf Suggested percentage P and its previous record as cluster-head. Nodes which were not cluster-head in previous 1/p rounds generate a number between 0 to 1 and if it is less then threshold T(n) then nodes become cluster-head. Threshold value is set through this formula.
\begin{eqnarray}
 T(n) = \left\{
  \begin{array}{l l}
    \frac {P}{1 - P*(rmod\frac{1}{P})} & \quad \textrm{if $n \in G $ }\\
    0 & \quad \textrm{otherwise}\\
  \end{array} \right.
\end{eqnarray}

Where G is set of nodes that have not been cluster-head in previous 1/p rounds, P= suggested percentage of cluster-head, r =is current round. The node becomes cluster-headin current round, it will be cluster-head after next 1/p rounds [1], [2], [3]. This indicates that every node will serve as a cluster-head equally and energy dissipation will be uniform throughout the network.  Elected cluster-head broadcasts its status using CSMA MAC protocol. Non-cluster-head node will select its cluster-head comparing RSSI of multiple cluster-head from where node received advertisements.Cluster-head will create TDMA schedule for its associated members in the cluster.\\
\begin{figure}[h]
  \includegraphics[scale=0.3]{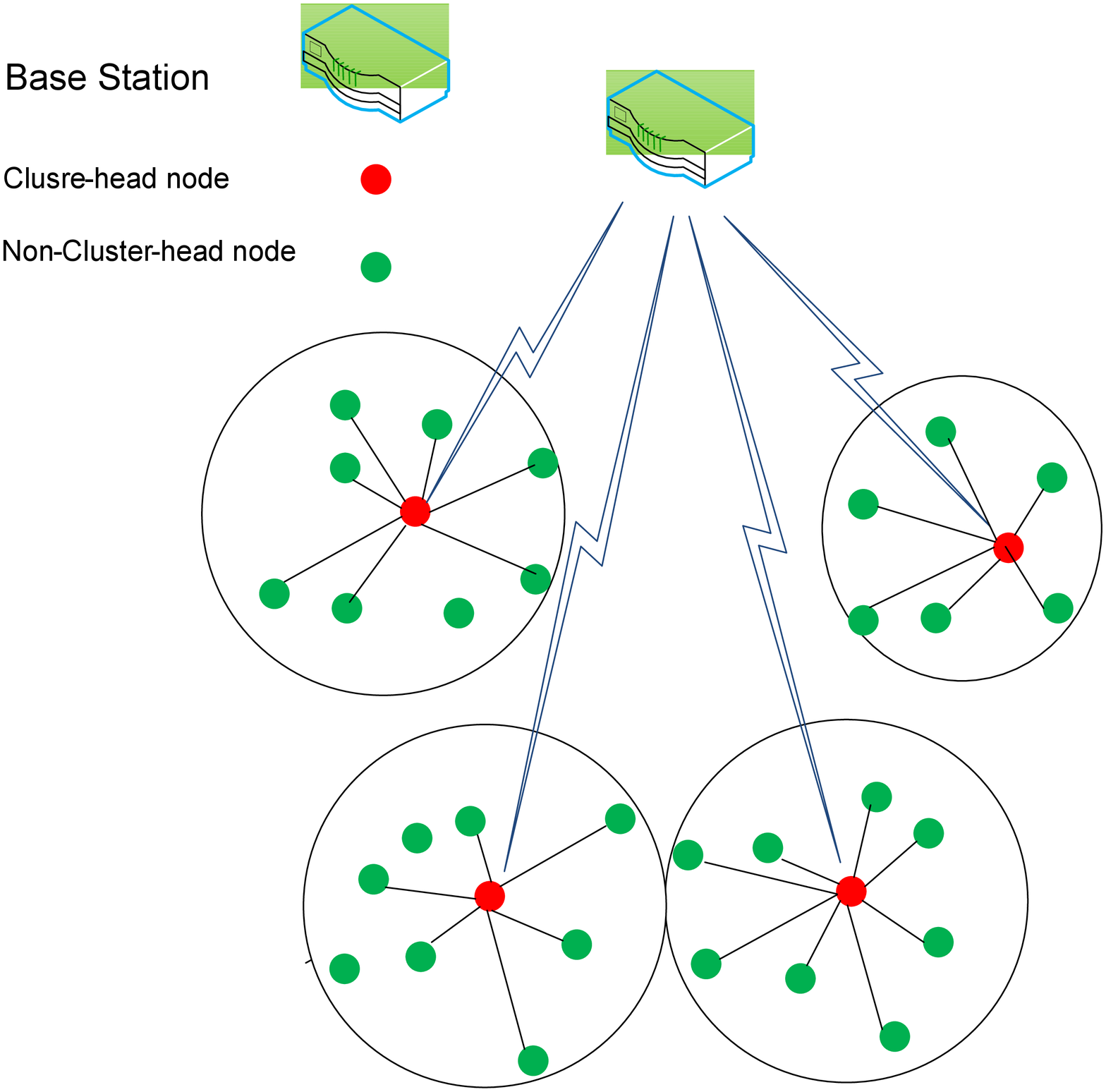}\\
  \caption{LEACH }\label{abc}
\end{figure}
In Steady state phase starts when clusters have been created. In this phase nodes communicate to cluster-head during allocated time slots otherwise nodes keep sleeping. Due to this attribute LEACH minimize energy dissipation and extend battery life of all individual nodes. When data from all nodes of cluster have been received to cluster-head.it will aggregate, compress and transmit to sink.  The steady state phase is longer than setup phase.\\

Figure 1. shows the LEACH basic communication hierarchy.
Energy (MTE) routing protocol. LEACH reduces this energy dissipation by following feature.\\
1.	Reducing the number of transmission to sink using cluster-head\\
2.	Reducing the date to be transmit through compression technique\\
3.	LEACH Increase the life time of all nodes through randomizes rotation being as cluster-head [1], [2], [3].\\
4.	LEACH allows non-cluster-head nodes to keep sleeping except specific time duration\\
5.	In LEACH routing protocol nodes die randomly and dynamic clustering enhance network lifetime\\
6.	LEACH routing protocol makes wireless sensor network scalable and robust\\
\subsection{Solar-aware Low Energy Adaptive Clustering Hierarchy(sLEACH)}
Energy harvesting is essential in some applications of wireless sensor network,especially when sensor nodes are placed in non-accessible areas like battlefield [9]. For such kind of applications solar-ware LEACH (sLEACH) has been proposed by authors [9]in which lifetime of the wireless sensor network has been improved through solar power. In sLEACH some nodes are facilitated by solar power and these nodes will act as cluster-heads mainly depending upon their solar status. Both LEACH and LEACH-C are extended by sLEACH.

\subsubsection{Solar-aware Centralized LEACH}
In solar-aware Centralized LEACH cluster head are selected by Base station with help of improved Central control algorithm. Base station normally select solar powered nodes as these have maximum residual energy. Authors improve the conventional cluster-head selecting algorithm used in LEACH-C [2], [3]. In sLEACH nodes transmit their solar status to base station along with energy and nodes with higher energy are selected as cluster-head. Performance of sensor network is increased when number of solar-aware nodes is increased. Sensor network lifetime also depends upon the sunDuration. It is the time when energy is harvested. If sunDuration is smaller cluster-head handover is also performed in sLEACH [9]. If node serving as cluster-head is running on battery and a node in cluster send data with flag, denoting that its solar power is increased this node will become cluster-head in place of its first serving cluster-head. This new cluster-head is selected during steady state phase that also enhance the lifetime of the network.

\subsubsection{Solar-aware Distributed LEACH}
In Solar-aware Distributed LEACH choosing preference of cluster-head is given to solar-driven nodes. Probability of solar-driven nodes is higher than battery-driven nodes. Equation 1 is needed to be change to increase the probability of solar-driven nodes. This can be achieved by multiplying a factor sf (n) to right side of the equation 1.
\begin{eqnarray}
 T(n)=sf(n) \times \frac{p}{1-(\frac{cHeads}{numNodes})}
\end{eqnarray}

Where sf (n) is equal to 4 for solar-driven nodes, sf (n) is equal to ¼ for battery driven nodes. P= is the percentage of optimal cluster-heads. The cHeads is number of cluster-heads since the start of last meta round. The numNodes is total number of nodes [8], [9].\\

\subsection{Multi-hop LEACH}
When the network diameter is increased beyond certain level, distance between cluster-head and base station is increased enormously. This scenario is not suitable for LEACH routing protocol [11] in which base station is at single-hop to cluster-head. In this case energy dissipation of cluster-head is not affordable. To address this problem Multi-hop LEACH  is proposed in [12].Multi-hop LEACH is another extension of LEACH routing protocol to increase energy efficiency of the wireless sensor network [11], [12], [13].Multi-hop LEACH is also complete distributed clustering based routing protocol. Like LEACH, in Multi-Hop LEACH some nodes elect themselves as cluster-heads and other nodes associate themselves with elected cluster-head to complete cluster formation in setup phase.

\begin{figure}[h]
  \includegraphics[scale=0.3]{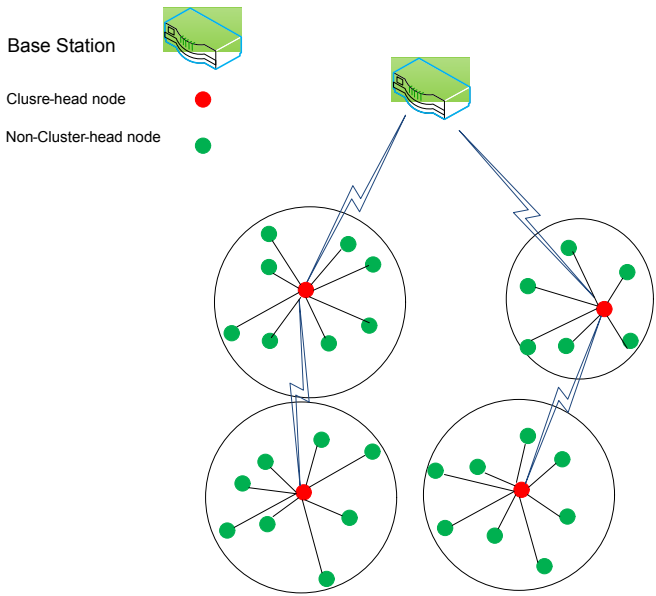}
  \caption{Multi-Hop LEACH}\label{ABC}
\end{figure}

In steady state phase cluster-head collect data from all nodes of its cluster and transmit data directly or through other cluster-head to Base station after aggregation. Multi-Hop LEACH allows two types of communication operations. These are inter-cluster communication and intra-cluster communication.
In Multi-hop inter-cluster communication, when whole network is divided into multiple clusters each cluster has one cluster-head. This cluster-head is responsible for communication for all nodes in the cluster. Cluster-head receive data from all nodes at single-hop and aggregate and transmit directly to sink or through intermediate cluster-head. In Multi-hop inter-cluster communication when distance between cluster-head and base station is large then cluster-head use intermediate cluster-head to communicate to base station.\\

Figure 2 describes Multi-Hop LEACH communication architecture.Randomized rotation of cluster-head is similar to LEACH. Multi-Hop LEACH selects best path with minimum hop-count between first cluster-head and base station.

\subsection{Mobile-LEACH (M-LEACH)}

LEACH considers all nodes are homogeneous with respect to energy which is not realistic approach. In particular round uneven nodes are attached to multiple Cluster-head; in this case cluster-head with large number of member ode will drain its energy as compare to cluster-head with smaller number of associated member nodes.  Furthermore mobility support is another issue with LEACH routing protocol, to mitigate these issues, M-LEACH is proposed in [16].\\

M-LEACH allows mobility of non-cluster-head nodes and cluster-head during the setup and steady state phase. M-LEACH also considers remaining energy of the node in selection of cluster-head. Some assumptions are also assumed in M-LEACH like other clustering routing protocols. Initially all nodes are homogeneous in sense of antenna gain, all nodes have their location information through GPS and Base station is considered fixed in M-LEACH. Distributed setup phase of LEACH is modified by M-LEACH in order to select suitable cluster-head. In M-LEACH cluster-heads are elected on the basis of attenuation model [17]. Optimum cluster-heads are selected to lessen the power of attenuation. Other criteria of cluster-head selection are mobility speed. Node with minimum mobility and lowest attenuation power is selected as cluster-head I M-LEACH. Then selected cluster-heads broadcast their status to all nodes in transmission range. Non-cluster-head nodes compute their willingness from multiple cluster-heads and select the cluster-head with maximum residual energy.\\
In steady state phase, if nodes move away from cluster-head or cluster-head moves away from its member nodes then other cluster-head becomes suitable for member nodes.It results into inefficient clustering formation.  To deal this problem M-LEACH provides handover mechanism for nodes to switch on to new cluster-head. When nodes decide to make handoff, send DIS-JOIN message to current cluster-head and also send JOIN -REQ to new cluster-head. After handoff occurring cluster-heads re- schedule the transmission pattern.

\section{Classification and comparison of LEACH and its modified routing protocols in wireless sensor networks}
Each routing protocol addresses specific problem and tries to enhance the conventional clustering routing protocol LEACH. Each routing protocol has some advantages and features. These routing protocol face some challenges like Cost of Clustering, Selection of Cluster-heads and Clusters, Synchronization, Data Aggregation, Repair Mechanisms, scalability, mobility, and initial energy level all nodes[14]. We compare above mention routing with respect to some very important performance parameters for wireless sensor network. These parameters are following.


Classification: The classifications routing protocol indicate that it is flat, location-based or hierarchal [15].

Mobility: it specifies that routing protocol is designed for fixed are mobile nodes. Scalability:  it how much routing protocol is scalable and can be efficient if the network density is increased.

Self-organization: it is very important for routing protocol to adopt the changes in network. Nodes configuration and re-configuration should be performed by routing protocol by self-organization at the time when nodes enter or leave the network [15].

Randomized Rotation of Cluster-head:  randomized Rotation of cluster-head is very necessary in order to drain the battery of all nodes equally [1].
Distributed clustering algorithm: cluster-heads are self-elected in distributed clustering algorithm also nodes select their cluster-head in distributed manner [1].\\
Centralized clustering algorithm: cluster-heads are selected by Base station by central control algorithm [3].

Single-hop or Multi-hop: it is also important feature of routing protocol. Single-hop is energy efficient if it is smaller area of network and multi-hop is better for denser network [11].

Energy Efficiency: it is the main concern of energy efficient routing protocol to maximize the life time of the network [1], [2], [4], [11], [15].\\
Resources awareness:  sensor network has limited resources like battery and sensing capability routing protocol should be well aware from the resources [8].\\
Data Aggregation:  in order to reduce the data amount to be transmit to Base station, Cluster-head perform data-aggregation in this way cluster-head transmission energy cost is reduce [1], [2].

Homogeneous: homogeneity of all nodes is considered in the all routing protocol which describe that initial energy level of all the nodes is similar.\\

\begin{table*}[t]
 \centering
  \caption{Performance comparison of hierarchical routing protocl}
    \begin{tabular}{|p{1.5cm}|p{1cm}|p{1cm}|p{1cm}|p{1cm}|p{1cm}|p{.7cm}|p{.7cm}|p{.7cm}|p{1cm}|p{1cm}|p{1cm}|p{1cm}|p{1cm}|}
  \hline

    Clustering Routing protocol & Classifi-cation & Mobility & Scalability & Self-organiza-tion & Randomi-
zed rota-tion
 & Distri-buted & Centra-lized  & Hop-count & Energy efficiency   & Resources awar-eness  & data aggregati-on & homogen-eous \\ \hline

    LEACH & Hierarchi-cal & fixed BS & imited & Yes   & yes   & yes   & No    & Single-hop & High  & Good  & yes   & yes \\ \hline

(sLEACH)
 & Hierarchi-cal & fixed BS & Good  & Yes   & yes   & yes   & yes   & Single-hop & Very High & Very Good & yes   & yes \\ \hline

(Multi-Hop LEACH)
 & Hierarchi-cal & fixed BS & Very Good & Yes   & yes   & yes   & No    & mult-hop & Very High & Very Good & yes   & yes \\ \hline
(M-LEACH)
& Hierarchi-cal & Mobile BS and nodes & Very Good & Yes   & yes   & yes   & No    & sinle-hop & Very High & Very Good & yes   & yes \\ \hline
    \end{tabular}%
  \label{tab:addlabel}%

\end{table*}%

Table.I shows the comparison LAECH,  sLEACH, M-LEACH and Multi-Hop LEACH. Performance comparison shows that these routing protocol are similar in many ways.  All routing protocol are hierarchal, homogeneous, having fixed BS despite M-LEACH, perform Data aggregation, self-organization and randomized rotation of CHs. LEACH, LEACH-SC, ELEACH, and Multi-Hop LEACH are use distributed algorithm for Cluster-head selection. LEACH-C uses central control Algorithm for cluster-head selection and sLEACH is designed for both centralized and distributed algorithm. LEACH, sLEACH and M-LEACH are routing protocol in which Base Station is at single-hop and in Multi-Hop LEACH Base station can be at multi-hop distance from the cluster-head. LEACH and M-LEACH allow limited scalability. sLEACH allows good scalability while Multi-Hop LEACH is providing maximum scalability feature due to multi-hop communication option for cluster-heads.

\section{Analytical comparison for Energy Efficiency of Routing Protocols}

For analytical comparison, it is essential to be aware from Radio model assumption adopted by energy efficient routing protocol. All energy efficient routing protocols proposed in previous research provide different assumptions about the radio distinctiveness. These different characteristics cause significant variation in energy efficiency of routing protocols. These assumptions differentiate energy dissipation to run transceiver and receiver circuitry per bit. Radio dissipates ?amp for transmit amplifier to attain suitable Eb/NO [1]. These are also multiple assumptions in selection of suitable ?amp.  Most acceptable value of these radio characteristics which is assumed by extensive research work is given in the table 2.\\

Transmitter and receiver Radio model is shown in figure 3.Mainly energy dissipation of a individual node depends upon the number of transmissions, number of  receiving, amount of data to be transmit and distance between transmitter and receiver. So first we describe the possible ways of energy consumption and then compare selected routing protocols and analyze how energy efficiency is enhanced through these routing protocol.

\begin{table}[t]
  \centering
  \caption{Radio Charactoristics}
    \begin{tabular}{|rrrr|rr|}
    \toprule
    Operation                                                           &       &       &       & Energy Dissipation &  \\
    \midrule
    Transmitter Electronics (EelectTx) &       &       &       & 50 nj/bit      &  \\
    Receiver Electronics (EelecRx)     &       &       &       & 50 nj/bit &  \\
    Transmit amplifier (Eamp)            &       &       &       & 100 pj/bit/m2 &  \\
    \bottomrule
    \end{tabular}%
  \label{tab:addlabel}%
\end{table}%

\begin{figure}[h]
  \includegraphics[scale=0.3]{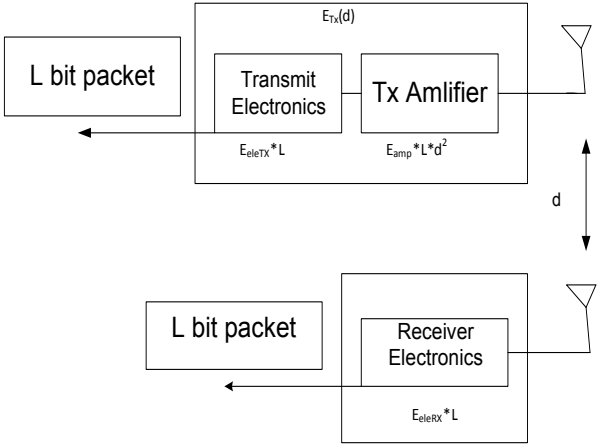}
  \caption{Radio Model }\label{abc}
\end{figure}

\subsection{Energy consumption  }

These are multiple source of energy consumption and every energy efficient routing protocol deals in different manner to reduce this energy consumption. This section provides  Mathematical analysis of these possible energy consumption sources. In mostly clustering routing protocol it is assume that nodes are uniformly distributed and there is free space communication model between all nodes and Base station.
as we know that in clustering routing protocols, duplex communication is needed some query-based application. During upward communication nodes send their data to cluster-head and cluster-head forwards aggregated data to Base station. so Energy consumption on cluster-head will be:
\begin{eqnarray}\nonumber
    E_{CH}=(\frac{n}{K}-1)(E_{eleRX}\times L{_{C}}+ \frac{n}{K}\times L{_{C}}E{_{AD}}+ \\ E{_{eleTX}}\times L_{A}+ E{_{amp}}\times L{_{A}}\times d_{to BS}^{2})
\end{eqnarray}

where n are all nodes in wireless sensor network, $E_{CH}$ is total upward communication energy consumption of Cluster-head, $(\frac{n}{K}-1)$ number of possible nodes in one cluster, K is possible number of clusters, $L{_{C}}$ is data of non-cluster-head nodes, $E{_{AD}}$ is energy cost for data aggregation, $L_{A}$ is aggregated data, $d_{to BS}^{2}$ is distance between cluster-head and Base station.\\

Energy consumption of single non-cluster-head node will be:
\begin{equation}\label{}
    E_{nonCH}=E{_{eleTX}}\times L_{C}+ E{_{amp}}\times L{_{C}}\times d_{to CH}^{2}
\end{equation}
where $d_{to CH}^{2}$ is distance between cluster-head and member node.
Energy consumption of all the nodes in one cluster will be:
\begin{equation}\label{}
    E_{nonCH}\times (\frac {n}{k}-1)= (\frac {n}{k}-1)( E{_{eleTX}}\times L_{C}+ E{_{amp}}\times L{_{C}}\times d_{to CH}^{2})
\end{equation}
so total  upward energy cost of single cluster will:
\begin{equation}\label{}
    E_{up}= E_{CH}+E_{nonCH}\times (\frac {n}{K}-1)
\end{equation}

When Base station has to get specific sensing information from nodes, in this case Base station send instructions to cluster-heads only and cluster-heads send these instructions to member nodes. in this process cluster-heads and non-cluster-heads pay energy cost. there is no issue of energy consumption on Base station so energy consumption of Base station is ignored. this downward energy cost is not consider in some cases , but it is certainly there. if Base station send instructions for all nodes to all cluster-head, the energy consumption on cluster-head will be:
\begin{eqnarray}\nonumber
    E_{CH}=(\frac{n}{K})E{_{eleRX}}\times L{_{BS}}+(\frac{n}{K}-1) \\ (E{_{eleTX}}+E{_{amp}}\times L{_{BS}}\times d_{to-nonCH})
\end{eqnarray}
when cluster-head transmits to  its member nodes, receiving nodes also consume energy and it will be equal to:

\begin{equation}\label{}
    E_{nonCH}(\frac{n}{K}-1)=(\frac{n}{K}-1)(E{_{eleRX}} \times L{_{BS}})
\end{equation}
Total downward energy consumption will be:
 \begin{equation}\label{}
    E{_{down}}=E_{nonCH}+E_{CH}
\end{equation}
So total estimated energy consumption for duplex communication of a single cluster will be:
\begin{equation}\label{}
    E{_{C}}=E{_{up}}+E{_{down}}
\end{equation}
From equation 10 total energy consumption of whole network can be also be estimated, and it will be:

\begin{equation}\label{}
    E{_{T}}=E{_{C}}\times K
\end{equation}
Clustering routing protocols for wireless sensor network also bear energy dissipation in setup phase. during cluster formation, energy is dissipated when  cluster-heads create TMDA schedule for member nodes. Every node keep sensing continuously it also pay energy cost. These kind energy costs are ignored incase of comparing clustering routing protocol.

\subsection{Energy Efficiency of Clustering Routing Protocols}
Hierarchical routing protocols we selected, compare their energy efficiency only with respect to upward transmission energy dissipation. in this scenario all nodes have to transmit their data to Base station through multiple cluster-heads. Distance between node and cluster-head play key rule in energy improvement. LEACH reduces energy dissipation over a factor of 7x and 8x reduction as compared to direct communication and a factor of4x and 8x compared to the minimum transmission [1]. this energy efficiency is due to reduction of number of direct transmissions because in LEACH only cluster-heads directly communicate to Base station and remaining nodes have to transmit to cluster-head which is at smaller distance.\\
The sLEACH is also provides better network life time as compare to LEACH.Because cluster-head selection is not uniform in sLEACH. In sLEACH solar-aware nodes are having more probability to be selected as cluster-heads as compare to battery-driven nodes. Multi-Hop LEACH is more energy efficient than LEACH [11].\\
Multi-Hop LEACH also provides better connectivity and successful data rate as compare to LEACH [12]. The reason behind this enhancement is multi-hop communication adopted by cluster-heads. As member nodes save energy by sending data to cluster-head in LEACH instead of Base station. Similarly in Multi-Hop LEACH cluster-head at longer distance from Base station transmit data to next cluster-head closer to Base station instead of direct transmission to Base station. Multi-Hop LEACH is more effective energy efficient routing protocol when network diameter is larger.Energy efficiency of multi-hop-LEACH can be better elaborate with the example of linear network shown in Figure4. In this network two cluster-heads A and B are communicating to Base station. Distance 'm' between Base station and two cluster-head is considered to be uniform.

\begin{figure}[h]
  \includegraphics[scale=0.3]{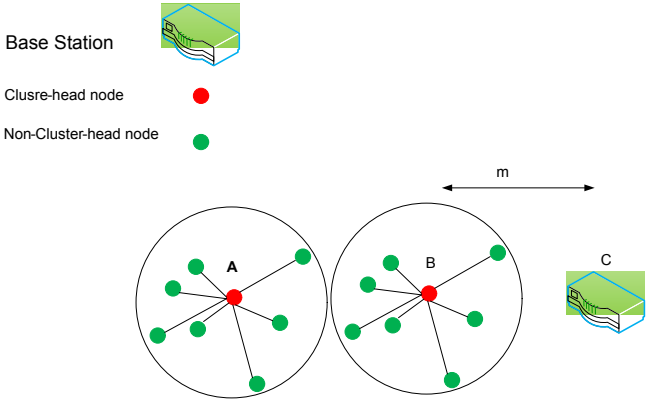}
  \caption{ Linear Network Model}\label{abc}
\end{figure}
In order to calculate the transmitting energy cost of cluster-heads A and B, which are directly transmitting to Base station will be:
\begin{eqnarray}\nonumber
    E_{dir}=E{_{eleTX}} \times L{_{A}}+ \epsilon {_{amp}}\times L{_{A}}\times 2m^{^{2}} +E{_{eleTX}} \times L{_{B}} \\ + \epsilon {_{amp}}\times L{_{B}}\times m^{^{2}}
\end{eqnarray}
Where $E_{dirAB}$ is total energy cost of cluster-heads A and B, $L{_{A}}$ is aggregated data transmitted by cluster-head A and $L{_{B}}$ is aggregated data transmitted by cluster-head B towards Base station and m is equal distance among cluster-heads and Base station. This happens in case of LEACH when every cluster-head has to communicate directly to Base station.\\

Similarly total transmitting energy cost can also be calculated when multi-hop communication is taking place. Multi-hop LEACH utilizes multi-hop communication.
In this linear network if cluster-head A transmits data to cluster-head B instead of Base station then cluster-head B has to transmit not only its own cluster's data but also has to transmit cluster-head B's data to Base station.
\begin{eqnarray}\label{}\nonumber
     E_{Multi-hop}=E{_{eleTX}} \times L{_{A}}+ \epsilon {_{amp}} \times L{_{A}}\times m^{^{2}}+E{_{eleRX}} \\ \times L{_{A}}+E{_{eleTX}} \times (L{_{B}}+L{_{B}})+ \epsilon {_{amp}}\times (L{_{B}}+L{_{B}})\times m^{^{2}}
\end{eqnarray}

Where $E_{Multi-AB}$ is total transmitting energy cost of both cluster-heads in case of multi-hop communication of Multi-hop LEACH. Cluster-head near base station has more traffic burden in case of M-LEACH. But cluster-head which is at longer distance from Base station has benefits because it has to transmit at small distance and increase its lifetime. M-LEACH is more efficient in case of large network diameter and LEACH is suitable when network diameter is small.

\section{Simulation Results and Analysis}
 we simulated LEACH, Multi-hop LEACH, M-LEACH and sLEACH-Centralized and sLEACH-Distributed to make efficient analysis. simulation parameters are shown in table 3. This simulation is implemented by using MATLAB. 100 nodes are scattered uniformly in region of 100m * 100m. During simulation of these routing protocols we adjusted the network topology according to realtime behavior of sensors nodes and also consider re-emargination ability of solar-driven sensors in sLECAH nodes to abstain more realistic simulation results.

\begin{table}[h]
  \centering
  \caption{Simulation Environment}
    \begin{tabular}{|rrrrrrrrrr|rr|}
    \toprule
    Parameter                                                           &       &       &       & value &  \\ \hline
    \midrule
    Network size        &       &       &       &            100m * 100m      &  \\\hline
    Initial Energy        &       &       &       &                .5 j      &  \\\hline
    p       &       &       &       &                      .1 j      &  \\\hline
    Data Aggregation Energy cost      &       &       &       &                      50pj/bit j      &  \\\hline
    number of nodes       &       &       &       &                     100      &  \\\hline
    packet size        &       &       &       &             200 bit       &  \\\hline
    Transmitter Electronics (EelectTx) &       &       &       & 50 nj/bit      &  \\\hline
    Receiver Electronics (EelecRx)     &       &       &       & 50 nj/bit &  \\\hline
    Transmit amplifier (Eamp)            &       &       &       & 100 pj/bit/m2 &  \\\hline

    \bottomrule
    \end{tabular}%
  \label{tab:addlabel}%
\end{table}%

 Figure.4 shows the network life time. in LEACH all nodes reach to death first and then M-LEACH, Multi-hop LEACH, sLEACH-Centralized and then sLEACH-distributed respectively. In solar-aware LEACH routing protocols nodes die after longest period of time because solar-awre nodes have ability to re-energize themselves for certain period. sLEACH has 300 $\%$ more network lifetime as compare to LEACH, because in sLEACH last node is dying after 4000 rounds. sLEACH efficiency can also be improved by adding more solar-driven nodes as compare to battery driven nodes. sLEACH-Distributed is slightly better then sLEACH-Centralized because in sLEACH-Distributed localized clustering formation is performed. M-LEACH has 30 $\%$ better network lifetime as compare to LEACH and last node reaches to death after 500 rounds. In Multi-hop LEACH routing protocol produces almost 40 $\%$ network life enhancement as compare to LEACH and it can be further improved if the network diameter is increases.

\begin{figure}[h]
  \includegraphics[scale=0.6]{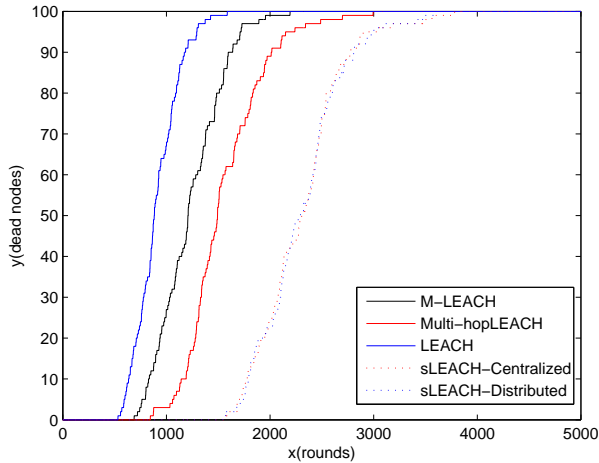}
  \caption{Dead Nodes }\label{abc}
\end{figure}

Figure.5 shows the number of allive nodes with respect to number of rounds for all selected routing protocols. Till 500 rounds all nodes are allive for every for every routing protocol. LEACH  has 30 $\%$, 40 $\%$ and 300 $\%$ less survival time as compare to M-LEACH, Multi-hop LEACH, and sLEACH respectively. Reason is similar as describe for figure 4.
\begin{figure}[h]
  \includegraphics[scale=0.6]{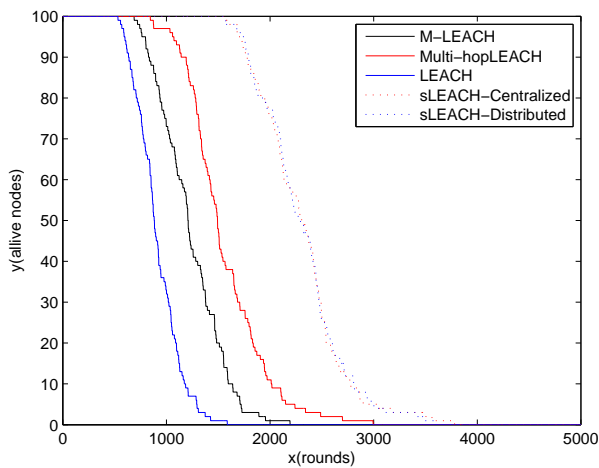}
  \caption{Allive Nodes }\label{abc}
\end{figure}

Quality of routing protocol is depending upon the date( actual data signal) successfully transfer to Base station. If data received by Base station is increasing it means quality of routing protocol is getting better and better.Figure.5 shows the quality analyzing clustering routing protocols.As cluster-heads are only responsible for aggregating and transmitting data to Base station, so routing protocol with optimum number of cluster-heads will be more efficient. Multi-hop LEACH has better quality than LEACH. Because in large network, cluster-head at the corner of the network have to transmit to next cluster-head towards Base station while in LEACH cluster-heads have to transmit directly to Base station at longer distance. that will result into poor signal strength and less successful data transmission. M-LEACH also provides better quality in dynamic topology of network. Comparatively sLEACH has maximum quality because in sLEACH cluster-heads are elected on the basis of solar property of nodes. Maximum cluster-heads in sLEACH are solar-driven nodes and these cluster-heads serve for longer period of time as compare to battery-driven cluster-heads in sLEACH. These solar-driven cluster-heads have enough energy to transmit at longer distance with acceptable signal strength that's why sLEACH has maximum quality of network. In sLEACH, sLEACH-Distributed has more quality as compare to sLEACH-Centralized. it is because of increasing probability of solar-driven nodes to be cluster-heads and its proved by equation 2 of this survey paper.
\begin{figure}[h]
  \includegraphics[scale=0.6]{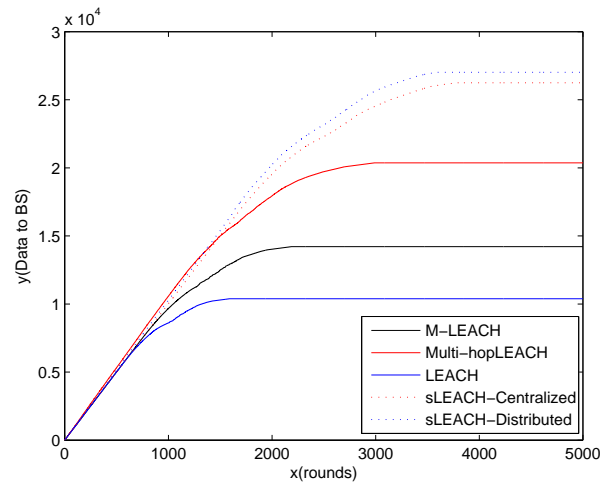}
  \caption{Number of Data signal received at the BS }\label{abc}
\end{figure}

As data to Base station is important factor for quality analysis of any routing protocol, similarly data(data signal) to cluster-head is also important. Figure.6 shows the data received by cluster-head. Results are similar as we computed from figure 5. however in this case sLEACH-Centralized is slightly better than sLEACH-Distributed. The reason behind this difference is central control algorithm used by sLEACH-Centralized. this central algorithm select cluster-head at suitable place for its member nodes in the cluster that's why sLEACH-Centralized has better quality than sLEACH-Distributed.
\begin{figure}[h]
  \includegraphics[scale=0.6]{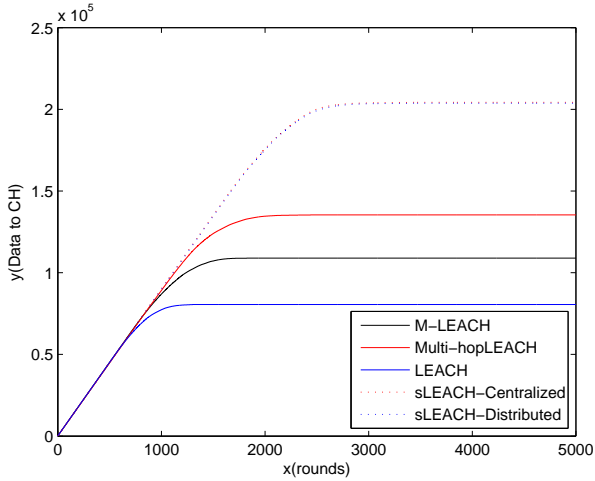}
  \caption{Number of Data signal received at the Cluster-head}\label{abc}
\end{figure}

AS LEACH, M-LEACH, Multi-hop LEACH and sLEACH-Distributed use distributed self-organization algorithm, because of this optimal number of cluster-heads are not guaranteed. Figure. 7 shows uncertain number of cluster-heads elected per rounds. Results shows that LEACH and M-LEACH show more uncertainty as compare to other routing protocols. sLEACH is slightly better incase of cluster-heads selection because criteria of randomized rotation of cluster-heads is modified from LEACH.
\\

\begin{figure}[h]
  \includegraphics[scale=0.6]{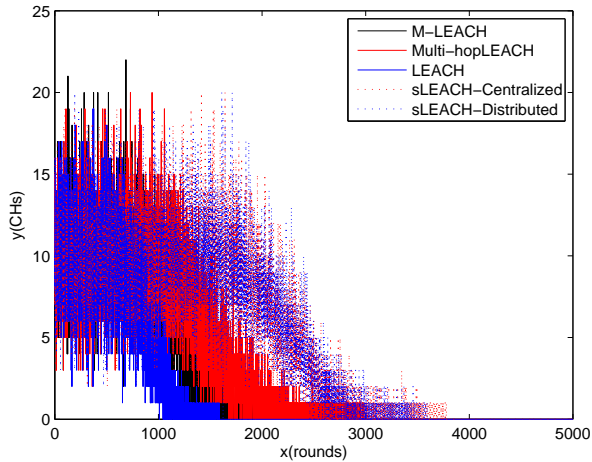}
  \caption{CHs per round }\label{abc}
\end{figure}

\section{Conclusion}
In this survey paper we have discussed LEACH, Multi-hop LEACH, M-LEACH Solar-aware LEACH hierarchical routing protocols for wireless sensor network. The main concern of this survey is to examine the energy efficiency and throughput enhancement of these routing protocols. We compare the lifetime and data delivery characteristics with the help of analytical comparison and also from our simulation results. Significant research work has been done in these different clustering routing protocols in order to increase the life time and data delivery features. Certainly further energy improvement is possible in future work especially in optimal guaranteed cluster-heads selection.

%

\end{document}